\documentstyle[preprint,prb,aps,epsf]{revtex}

\begin{document}
\draft
\preprint{}
\sloppy
\bibliographystyle{plain}

\title{\bf Enhanced vortex damping by eddy currents in superconductor-semiconductor hybrids} 

\author{M.~Danckwerts$^1$, A.R.~Go\~{n}i$^1$,
  C.~Thomsen$^1$, K.~Eberl$^2$, and A.G.~Rojo$^3$} 

\address{$^1$Institut f\"{u}r Festk\"{o}rperphysik, Technische 
Universit\"at Berlin, Hardenbergstr. 36, 10623 Berlin, Germany}

\address{$^2$Max-Planck-Institut f\"{u}r Festk\"{o}rperforschung, 
Heisenbergstr. 1, 70569 Stuttgart, Germany}

\address{$^3$Dept.~of Physics, University of Michigan, Ann Arbor,
MI 48109, USA}

\maketitle

\begin{abstract}
\noindent
{\small 
An enhancement of vortex-motion damping in thin Pb/In superconducting
films is obtained through coupling to an adjacent two-dimensional
electron gas formed in a modulation-doped GaAs/AlGaAs heterostructure.
This effect is observed by monitoring the power dissipation 
at the superconductor in the vortex state while increasing the
density of the electron gas using a gate voltage. Quantitative agreement
is found with calculations based on a viscous model of vortex damping
which considers generation of eddy currents in the electron gas by
moving flux lines. In the regime of filamentary and channel vortex
flow, eddy-current damping leads to striking dissipation breakdown
due to stopping of entire vortex channels.}
\end{abstract}


Superconductor-semiconductor hybrid structures are 
emerging as key devices in the search for new physical phenomena 
resulting from interactions between two systems with dissimilar 
electronic properties \cite{schef96a}. In particular, Josephson-type 
junctions with Nb electrodes coupled by a two-dimensional electron 
gas (2DEG) in InAs layers exhibit phase-sensitive transport due to 
Andreev reflections of quasi particles at the interfaces between normal 
metal and superconductor \cite{nguye92a,dimou95a,weesx96a}. Other 
experiments concentrate on commensurability
and interference effects on electron ballistic transport in the 2DEG,
which occur when a perpendicular magnetic
field is spacially modulated by the vortices of an adjacent 
superconducting film. In this case, a pronounced suppression of
the Hall effect was observed and ascribed to electron diffraction
by flux quanta \cite{geimx92a}.

Few investigations are concerned with the influence of a normal metal
on the vortex dynamics under a transport current, although the devices
were intentionally designed to have no viscous coupling \cite{kruit91a}. 
One can think of our hybrid system as a modified
Giaever's dc transformer \cite{giaev65a}, in which one of the 
superconducting films has been replaced by a 2D electron gas.
Under the action of a Lorentz force the vortices move at constant velocity
due to viscous damping.
For an isolated superconductor, this damping originates from the voltage
induced across the normal core of each moving vortex.
By bringing a highly mobile electron gas close enough to the 
superconducting film, i.e.~at a distance of the order of the London 
penetration depth, an additional dissipation mechanism is 
introduced through magnetic coupling resulting in an increase of viscosity. 
An interesting issue is to what extent this would affect
filamentary and channel vortex flow, for which dissipation jumps 
are observed in the current-voltage curves 
\cite{danck99a,helle96a,gronb96a}. 
The study of vortex damping in hybrids may provide
further insight into vortex-vortex interactions and pinning effects.  

This Letter reports the first observation of damping enhancement 
for vortex motion due to the presence of a high-mobility electron gas 
in superconductor-semiconductor hybrids. The samples used in our
experiments consist of thin Pb/In
films evaporated on top of modulation-doped GaAs/AlGaAs
heterostructures. The evidence is
found in the decrease of dissipation voltage 
measured at the superconducting film due to a higher
viscosity for vortex flow in the hybrid system, as compared to the case
without the 2DEG beneath. We vary the normal metal conductivity 
by increasing the carrier density using a gate voltage applied 
between the 2DEG and a back
contact. Our results are in quantitative agreement with the 
predictions of a model which accounts for the generation of eddy
currents in the electron gas by flowing vortices. 
For an estimated increase in electron density of up to 20\% the 
relative change observed in dissipation voltage lies in the one-percent
range. For the currents and magnetic fields at which filamentary vortex
flow occurs, however, striking dissipation reductions 
are readily achieved.  


The semiconductor component of the hybrid system is either a 
25 nm wide GaAs/AlGaAs single quantum well (SQW) or a
single heterointerface (SHI) structure.
A two-dimensional electron gas is realized
by modulation doping and is buried at a distance 
$D=$~75 nm and 50 nm from the surface for the SQW and SHI structure,
respectively \cite{ernst94a}. The nominal mobilities and carrier densities
of both samples at 4.2 K and under illumination are 
$\mu\simeq 8\times 10^{5}\, \mathrm{cm}^{2}/\mathrm{Vs}$ and 
$n\simeq 5.6\times10^{11} \, \mathrm{cm}^{-2}$.
The electron gas is contacted from the surface by In alloying in order 
to apply a gate voltage $U_{g}$ between it and a metallic back contact. 
The variation of the 2D density was examined previously in
photoluminescence experiments \cite{gonix99a}. A linear increase in the
carrier density $n_{\mathrm{2D}}$ between its nominal value and 
at most $\sim 6.5\times10^{11} \, \mathrm{cm}^{-2}$ 
can be achieved by applying a gate voltage
between 0 and 200 V. Hence the 
estimated maximum possible increase of density is less than 20\%. 
Superconducting films of Pb with nominally 14
at.\% In were evaporated on the semiconductor surface
($4 \times 4$ mm$^{2}$) with 
film thicknesses $d$ ranging from 60 to 300 nm, as determined using
atomic-force microscopy. The superconducting transition in zero
field occurs at 7.2 K. For transport experiments, Au leads were pressed 
against the superconductor film. Current-voltage measurements were
performed with standard four-terminal configuration 
using dc currents up to 1.5 A. Experiments were carried out at
4.2 K and low perpendicular magnetic fields $B <$ 0.2 T.


To model the coupling between vortex lattice and electron gas in our
hybrid samples we consider the effect on the normal metal of 
the magnetic field $B$ of a moving vortex with speed $v$. 
The experimental situation is schematically shown in the inset to
Fig.~1. Flowing vortices induce an electric field in
the 2DEG that generates eddy currents leading to an additional
dissipation which, in turn, forces the fluxoids to slow down. 
In the limit $D\ll\lambda^{2}/d$ of small 
superconductor-2DEG distances as compared to the effective 
London penetration depth, the magnetic field of a vortex can be 
approximated as
${\mathbf{B}}\sim (\Phi_{0}/2\pi\lambda^{2})K_{0}(r/\lambda)\hat{z}$,  
where $\Phi_{0} = h/2e$ is the flux quantum and $K_{0}$ the
zeroth-order Bessel function of imaginary argument \cite{clemx80a}.
For our samples the effective penetration depth 
is five to eight times $D$ \cite{danck99a}.
The vector potential in the plane of the 2DEG can be written as
\begin{equation}
  \label{eq:vec-pot}
  {\mathbf{A}}(\rho) =
  \frac{\Phi_{0}}{2\pi \rho} {F(\rho/\lambda)} \hat{e}_{\varphi},
\end{equation}
\noindent
where $\rho$ is the polar radius from the vortex core,
 and $F(x)=\int_0^x  dy\, yK_0(y)$.

The time-varying vector potential 
produced by a flowing vortex induces an electric field 
${\mathbf{E}} = v \,
\frac{\partial}{\partial x}{\mathbf{A}}(x-vt,y)$, which causes
joule dissipation in the 2D gas. The energy loss per unit time 
is calculated according to
\begin{equation}
  \label{eq:p-diss}
  \frac{d\varepsilon}{dt} =  \sigma_{\mathrm{2DEG}}
  v^{2} \, \int d^{2}x \left[\frac{\partial}{\partial
  x}{\mathbf{A}}(\rho/\lambda)\right]^{2} 
  \simeq \frac{\Phi_{0}^{2}}{2\pi\lambda^{2}}
   \sigma_{\mathrm{2DEG}} v^{2}\equiv  \eta_{\mathrm{2DEG}} \, v^{2},
\end{equation}
\noindent
where $\sigma_{\mathrm{2DEG}}$ is the conductivity of the 2D electron
gas and the dimensionless integrals are assumed to be of the
order of one. 

We arrive at the general result that eddy current generation in
the hybrid system manifests itself in a contribution to the viscosity  
$\eta_{\mathrm{2DEG}}$
describing the enhanced damping of vortex motion. 
This effect can be observed if $\eta_{\mathrm{2DEG}}$
is comparable to the viscosity of type-II superconducting material,  
$\eta_{\mathrm{SC}} = \sigma_{n} d\, \Phi_{0}^{2}/2\pi a^{2}$
\cite{tinkh96a}, where $\sigma_{n}$ is the normal state conductivity 
of the superconductor and $a$ the vortex core radius. The electron 
gas acts as a shunt conductor, thus increasing the system viscosity 
$\eta_{\mathrm{tot}}=\eta_{\mathrm{SC}}+\eta_{\mathrm{2DEG}}$ 
by a factor 
\begin{equation}
  \label{eta/eta}
1+\frac{\eta_{\mathrm{2DEG}}}{\eta_{\mathrm{SC}}} =
1+\left(\frac{a}{\lambda}\right)^{2} \,
  \frac{\sigma_{\mathrm{2DEG}}}{\sigma_{n}d}.
\end{equation}
\noindent
Hence the dissipation voltage under a transport current is 
$U_{d}\propto v$ \cite{tinkh96a} and the vortex velocity is
determined by the balance between the driving force $jB$ and the viscous
drag $\eta v$. Eddy-current damping grows in proportion to
$\sigma_{\mathrm{2DEG}}$, i.e. to the carrier density
$n_{\mathrm{2D}}$ of the electron gas. 

The dissipation due to flowing vortices in the PbIn superconducting film
is reduced by increasing the charge density in the
neighboring electron gas. Figure 1 shows a typical 
dissipation voltage ($U_d$) versus gate bias ($U_g$) curve of a 
PbIn/SHI hybrid sample measured at a constant transport current 
of 600 mA. Sweeping $U_g$ from 0 to 170 V causes a linear decrease in 
dissipation of $\Delta U_{d}=0.022$ mV, i.e. $\sim0.1$ \% of the 
initial value. Since optical and transport experiments could not be 
carried out simultaneously, we give here the gate voltage as a 
measure of the electron density which we assume to vary linearly
with $U_{g}$, as inferred from optical measurements \cite{gonix99a}.
We emphasize that in spite of the relatively small change in 2DEG
density the coupling within the hybrid appears to be effective 
enough to show up in its dissipative behavior. We also notice 
that during the experiment no leakage current between superconductor
and electron gas was ever detected, thus we rule out any
spurious bias to be at the origin of the observed effect.

We interpret the dissipation change in the superconductor with
rising charge density in the 2DEG as the
effect of eddy currents in the electron gas which slow down the vortices
causing the voltage across the superconductor to decrease.
Using Eq. (\ref{eta/eta}) we can now estimate the magnitude of
this effect. Taking $a\approx\lambda$, $d\approx 100$ nm,
$\sigma_{n}\approx1.4\times10^{5}\,\Omega^{-1}\mathrm{cm}^{-1}$ as
obtained from resistance measurements, and
$\sigma_{\mathrm{2DEG}}\approx 0.08\,\Omega^{-1}$ with
the 2DEG parameters given above, this yields
$\eta_{\mathrm{2DEG}}/\eta_{\mathrm{SC}} \approx 5 \%$.
This change in dissipation voltage corresponds to the 
difference between having and not having the electron gas next to the 
superconductor. In our experiments, however, we 
start from a finite density and
produce a variation of about 10\%. Thus, the calculated 
dissipation change is around 0.5\%,
in very good agreement with the experimental results.

Eddy-current damping effects are much more pronounced
in the regime of filamentary and channel vortex flow, for they 
can lead to a striking fall of dissipation voltage by more 
than one order of magnitude, as shown in Fig.~2. Here, $U_d$ was 
measured at a current close to the repinning transition of a large 
vortex channel. The inset to Fig.~2 displays the $IV$ characteristic 
of the superconducting film measured for a field of 53 mT and 
at 4.2 K but without gate bias. The abrupt jumps and large 
hysteresis apparent in the $IV$ curve are the signature of 
channel vortex flow \cite{danck99a}. As indicated by the arrow,
the point where the measurement of dissipation versus gate
voltage was carried out is close above the critical current 
at which the downward jump occurs. In this case,
the vortex speed is just high enough for the channel to keep
flowing. Energy loss due to eddy currents slows the vortices
further down, so that the whole channel will eventually be repinned.
When the vortex channel stops, dissipation suddenly drops.
This process is irreversible since, as can
be seen in Fig.~2, dissipation does not resume to its initial
value when the gate voltage is swept back to zero. 
This effect is highly reproducible even after heating 
the sample over $T_{c}$ and re-cooling.

The study of the dependence on gate voltage of the 
current values at which dissipation jumps occur 
provides further information about the nature of
filamentary vortex flow and the role played by pinning. 
Figure 3 shows one example in which the currents for the upward and 
downward voltage jumps are plotted as a function of gate bias (the
corresponding $IV$ curve is displayed in the inset to Fig. 3).
The jump-up current is independent of the 2D density indicating
that the pinning strength is not appreciably affected by the presence 
of the electron gas. In contrast, the jump-down current increases with $U_g$. 
With increasing viscosity the vortices of a moving channel slow down such 
that its repinning occurs at larger values of the transport current. 
This is interpreted as additional evidence of a repinning force which 
depends on vortex velocity. A similar increase in jump-down current 
is observed for thin PbIn films on glass by decreasing the external 
magnetic field \cite{danck99a}.

The influence of magnetic field homogeneity on eddy-current
damping is revealed by the percentage change in dissipation 
as a function of magnetic field at constant transport current. 
As a measure of the damping strength, the maximum
dissipation change $\Delta U_{d}$ for a gate voltage interval 
of $\Delta U_{g}=170$ V is normalized by the  
dissipation voltage $U_{d0}$ at zero bias. The 
corresponding values measured at 500 mA and 1000 mA are shown in Fig.~4
as a function of the external magnetic field $B$.
At $B\le 30$ mT damping causes 0.1--0.2 \%
change in dissipation. At 40 mT, the 
data for both 0.5 A and 1 A display a sharp maximum and become very
small at fields larger than 50 mT, where the resistance of
the film is close to normal but the transport behavior is
characterized by massive vortex flow.

The weakening of the effect of eddy-current damping on dissipation 
with increasing magnetic field (solid
line in Fig.~4) can be explained as due to the growing
homogeneity of the field pattern of the vortices while
approaching the upper critical field $B_{c2}$. 
At low magnetic fields, i.e.~low vortex density, 
the field distribution is very inhomogeneous, since $B$ has a maximum
at the vortex cores dropping to zero between them.
Thus, $\dot{B}$ is large and damping is efficient.
When vortices start to overlap, the lateral modulation of 
magnetic field in the plane of the 2DEG is continuously reduced. 
As a consequence, eddy currents as well as damping effects are weak.
In contrast, the peak at 40 mT is associated with the enhancement 
of the effects due to eddy-current damping in the regime dominated
by filamentary vortex flow.
Here the dissipation change is greatly enhanced by fluctuations in the
number of moving vortices contributing to dissipation due to the 
depinning and repinning of a large number of small vortex filaments
in quick succession.


In summary, we have observed significant additional damping of vortex
motion in superconductor-semiconductor hybrid
systems. A theoretical model is used to calculate the damping effect
from eddy currents generated in the 2D electron gas showing quantitative
agreement with the experiment. Under conditions of filamentary
vortex flow, the energy loss 
due to eddy currents leads to the stopping of entire channels, such that
power dissipation in a hybrid device can be switched off by slightly
increasing the electron density. At large fields damping is
weak due to the vanishing lateral field modulation in the plane of the
2DEG. We point out that, although our
observations can be explained within the framework of
classical electrodynamics, novel effects due to quantization of the 
electron gas conductivity are anticipated to occur
for the conditions of the experiments. Our results provide further 
insight into the issue of vortex dynamics with dissipation and open 
up a new class of devices for the study of
correlations between adjacent non-tunneling systems with dissimilar
electronic and magnetic properties.

A.G.R. acknowledges useful conversations with John Clem and Cagliyan
Kurdak and partial support from the National Science Foundation.

\newpage

\figure{{\bf Fig.~1.} 
Dissipation voltage as a function of gate bias of a PbIn/SHI hybrid 
structure in a magnetic field of 53 mT and at 4.2 K. The transport 
current was 600 mA. The inset shows a sketch of the hybrid sample.
$I$ is the transport current. The arrow represents a
flux line with magnetic field $B$ moving at a speed $v$.}

\figure{{\bf Fig.~2.} 
Dissipation voltage as a function of gate bias of a PbIn/SHI hybrid 
structure in a perpendicular magnetic field of 53 mT for 500 mA
current and at 4.2 K. The inset shows the corresponding current-voltage 
characteristic. The point where the measurement of dissipation versus 
gate voltage was taken is marked with an arrow.}

\figure{{\bf Fig.~3.} 
Dependence on gate voltage of the critical current for the upward 
(up triangles) and the downward voltage jump (down triangles) for
a PbIn/SQW hybrid at 4.2 K and at a magnetic field of 71 mT within the
region of channel vortex flow. 
The inset shows the corresponding $IV$ characteristic.}

\figure{{\bf Fig.~4.} 
Dissipation percentage change versus magnetic field for a PbIn/SQW
hybrid sample 
at 4.2 K. Data were taken at 500 mA (solid symbols) and 1000 mA (open
symbols). The solid line is a guide to the eye. The shaded area 
indicates the field range within which voltage jumps are observed 
for the used currents.}

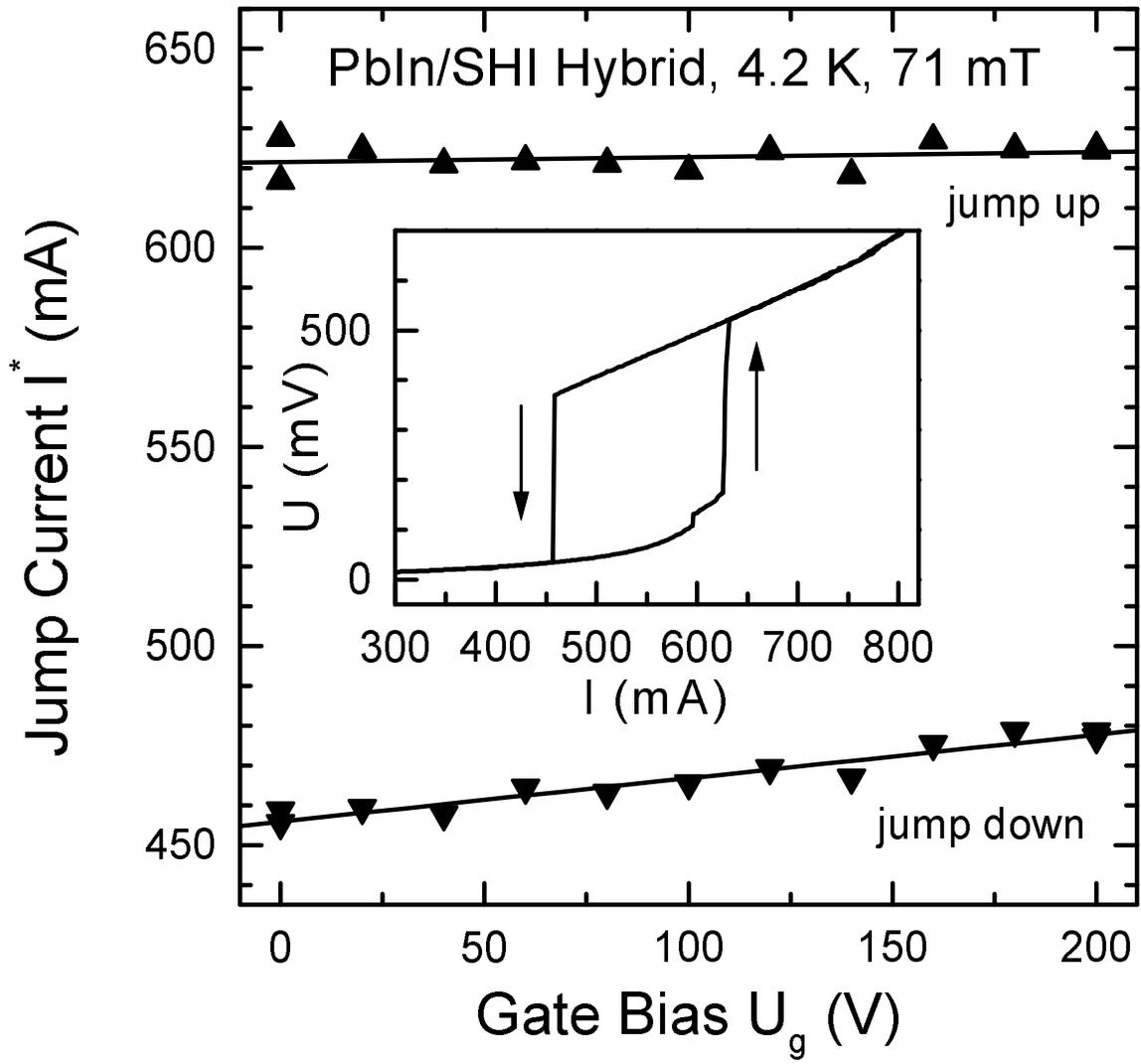
\end{document}